\documentclass[twocolumn,english,aps,pra,showpacs]{revtex4-1}
\usepackage[T1]{fontenc}
\usepackage[latin9]{inputenc}
\usepackage{geometry}
\usepackage{color}
\usepackage{amsbsy}
\usepackage{amstext}
\usepackage{amssymb}
\usepackage{graphicx}
\usepackage{esint}

\makeatletter
\usepackage {url}

\makeatother

\usepackage{babel}
\begin{document}

\title{Quantum signature for laser-driven correlated excitation of Rydberg
atoms}

\author{Huaizhi Wu,$^{1,2}$ Yong Li,$^{3,4}$ Zhen-Biao Yang,$^{1,2}$ and Shi-Biao
Zheng$^{1,2}$}

\address{$^{1}$Department of Physics, Fuzhou University, Fuzhou 350116, People\textquoteright s
Republic of China, $^{2}$Fujian Key Laboratory of Quantum Information and Quantum Optics, Fuzhou University, Fuzhou 350116, People's Republic of China, $^{3}$Beijing Computational Science Research Center,
Beijing 100193, People\textquoteright s Republic of China,$^{4}$Synergetic Innovation Center of Quantum Information and Quantum Physics, University of Science and Technology of China, Hefei, Anhui 230026, China}
\begin{abstract}
The excitation dynamics of a laser-driven Rydberg atom system exhibits
cooperative effect due to the interatomic Rydberg-Rydberg interaction,
but the large many-body system with inhomogeneous Rydberg coupling
is hard to be exactly solved or numerically studied by density matrix
equations. In this paper, we find that the laser-driven Rydberg atom
system with most of the atoms being in the ground state can be described
by a simplified interaction model resembling the optical Kerr effect
if the distance-dependent Rydberg-Rydberg interaction is replaced
by an infinite-range coupling. We can then quantitatively study the
effect of the quantum fluctuations on the Rydberg excitation with
the interatomic correlation involved and analytically calculate the
statistical characteristics of the excitation dynamics in the steady
state, revealing the quantum signature of the driven-dissipative Rydberg
atom system. The results obtained here will be of great interest for
other spin-1/2 systems with spin-spin coupling.
\end{abstract}
\maketitle

\section{Introduction }

A laser-driven Rydberg gas in which the Rydberg excited atoms experience
the long-ranged dipole-dipole or van der Waals interaction potential
can exhibit atomic many-body correlations that are of great interest
for future applications in quantum information processing \cite{Saffman_RMP2010}
and quantum nonlinear optics \cite{Chang2014}. While much attention
has been recently devoted to the enhancement of optical nonlinearity
by mapping the Rydberg-Rydberg correlation onto optical field \cite{Pritchard_PRL2010,Sevincli2011,Petrosyan_PRL2011,Peyronel2012,Parigi_PRL2012,Yan_PRA2012,Gorshkov_PRL2013,PhysRevLett.110.103001,He2014a,Li2014,Maghrebi2015a}
and to the nonequilibrium quantum phenomena \cite{PhysRevLett.101.250601,Weimer2010,PhysRevA.84.031402,PhysRevA.85.065401,PhysRevLett.111.215305,Marcuzzi2014a,Hoening2014a}
and the preparation of Rydberg crystals \cite{Glaetzle_PRA2012,Honing2013,Vermersch2015}
by modulating the driven-dissipative dynamics, the quantitative understanding
of the excitation process for the Rydberg atom system itself is of
particular interest on the other hand and is not yet clear. 

There has been a lot of theoretical work focusing on the excitation
dynamics in both the coherent \cite{PhysRevLett.98.023002,Ates2007,Mayle2011,Ates2012a,Petrosyan2013,Garttner2013a,PhysRevA.88.043436,Sanders2015}
and dissipative regimes \cite{Ates2012,Lee2012,PhysRevLett.108.023602,Hu2013,Schonleber2014b,Garttner2014a,Olmos2014,Weimer2015a,Schempp2015}.
Generally, the system was analyzed via a mean-field treatment disregarding
the interatomic correlation \cite{Ates2012,Marcuzzi2014a,Olmos2014,Vermersch2015,PhysRevLett.111.215305},
a variational approach allowing for an approximation of the true steady
state \cite{Weimer2015a} or a perturbation theory up to fourth order
\cite{Lee2012,Garttner2014a,Schempp2015}. Alternatively, the incoherent
dynamics can be numerically simulated on account of strong dissipation
via the rate equation \cite{Ates2007,Schonleber2014b}, and weak decay
via the density-matrix master equation \cite{Petrosyan2013} (or equivalently
the wave-function Monte Carlo approach \cite{Lee2012,PhysRevLett.108.023602,Hu2013,Olmos2014,Schonleber2014b})\textcolor{black}{{}
and the density-matrix renormalization-group \cite{Honing2013}} that
are normally useful for the lattice geometrics consisting of several
tens of atoms.

The excitation dynamics under resonant and off-resonant driving exhibits
different counting statistics of the Rydberg atom number. While the
resonant driving leads to reduced number fluctuations of the Rydberg
excitations \cite{PhysRevLett.109.053002,Hofmann_PRL2013,PhysRevA.88.043436},
the off-resonant coupling offers richer physics such as the experimentally
observed optical bistability \cite{PhysRevLett.111.113901}, Rydberg
aggregates \cite{Schempp2014,Urvoy2014a}, bimodal counting distribution
\cite{Malossi2014}, and kinetic constraints \cite{Valado2015}. In
addition, of special interest for the experimental analysis is the
formation mechanism of the collective many-body states since the statistical
characteristics \cite{Schempp2014,Malossi2014} and direct imaging
\cite{Gunter2012,Schwarzkopf2011_Imaging_detection} are challenging
for distinguishing sequential and simultaneous excitation process
of the Rydberg atoms. These have also been theoretically studied via
the analytical model with generalized Dicke states \cite{PhysRevLett.109.053002}
or Monte Carlo simulation \cite{Schempp2014}. 

In this paper, we study the excitation dynamics of a laser driven-dissipative
Rydberg atom system in the Holstein-Primakoff regime, i.e. the majority
of atoms remain in the ground state
\cite{Schempp2014,Malossi2014}. To give an intuitive understanding,
we make use of a simplified picture for the many-body system as in
ref. \cite{PhysRevLett.108.023602}, where the characteristic $C_{\mu}/R^{\mu}$
($\mu=3,6$) dependent interaction for atom pairs with the separation
$R$ is replaced by an infinite-range coupling (or average pair interaction).
It allows us to regard the system as a nonlinear optical polarizability
model \cite{Drummond_JPA} and quantitatively study the effect of
quantum correlations on the excitation dynamics in the steady state.
We find from the linearized calculation that the quantum fluctuations
to the first order will enhance the Rydberg population \cite{PhysRevLett.108.023602,Garttner2014a}
and modulate the fluctuation of the collective excitation number.
Moreover, the nonclassical effects like pair excitation of Rydberg
atoms can be clearly revealed by the full quantum results after comparison
with the classical steady-state solution. The limitations for this simplified model are also
discussed. Our finding not only connects to the recent experimental
observations for the many-body system with Rydberg-Rydberg coupling
\cite{Labuhn_Nature2015}, but also relates to the trapped spin-$1/2$
ions system where the spin-spin couplings are mediated by the motional
degrees of freedom \cite{Britton_Nature2012}.

The paper is organized as follows. In Sec. II, we introduce our model
for the interacting Rydberg atoms in the Holstein-Primakoff regime.
In Sec. III, we numerically solve the classical Langevin equation
of motion for the Rydberg atom system to study the bistability of
Rydberg population. In section IV, we discuss the effect of quantum
fluctuations on the Rydberg excitation dynamics in the linearized
regime. In section V, we give the exact steady-state solution of the
Fokker-Planck equation for describing the full quantum dynamics. Section
VI finally contains a summary of our results and an open discussion of the limitations for the model.

\section{Theoretical Model}

Consider a system of $N$ atoms ($N\gg1$) that are excited from the
ground state $|g_{j}\rangle$ to a Rydberg state $|e_{j}\rangle$
by a continuous and spatially uniform laser beam with the detuning
$\Delta$ from atomic resonance and the Rabi frequency $\Omega$ (assumed
to be real). By including the all-to-all interatomic coupling (or
average pair interaction) $\chi$, the Hamiltonian in the interaction
picture and rotating-wave approximation reads ($\hbar=1$)\textcolor{blue}{{}
}\cite{PhysRevLett.108.023602}

\begin{eqnarray}
H & = & \sum_{j=1}^{N}\left[-\Delta|e_{j}\rangle\langle e_{j}|+\frac{\Omega}{2}(|e_{j}\rangle\langle g_{j}|+|g_{j}\rangle\langle e_{j}|)\right]\nonumber \\
 &  & +\chi\sum_{j<k}|e_{j}\rangle\langle e_{j}|\otimes|e_{k}\rangle\langle e_{k}|.\label{eq:H_MeanF}
\end{eqnarray}

\begin{figure}
\includegraphics[width=1\columnwidth]{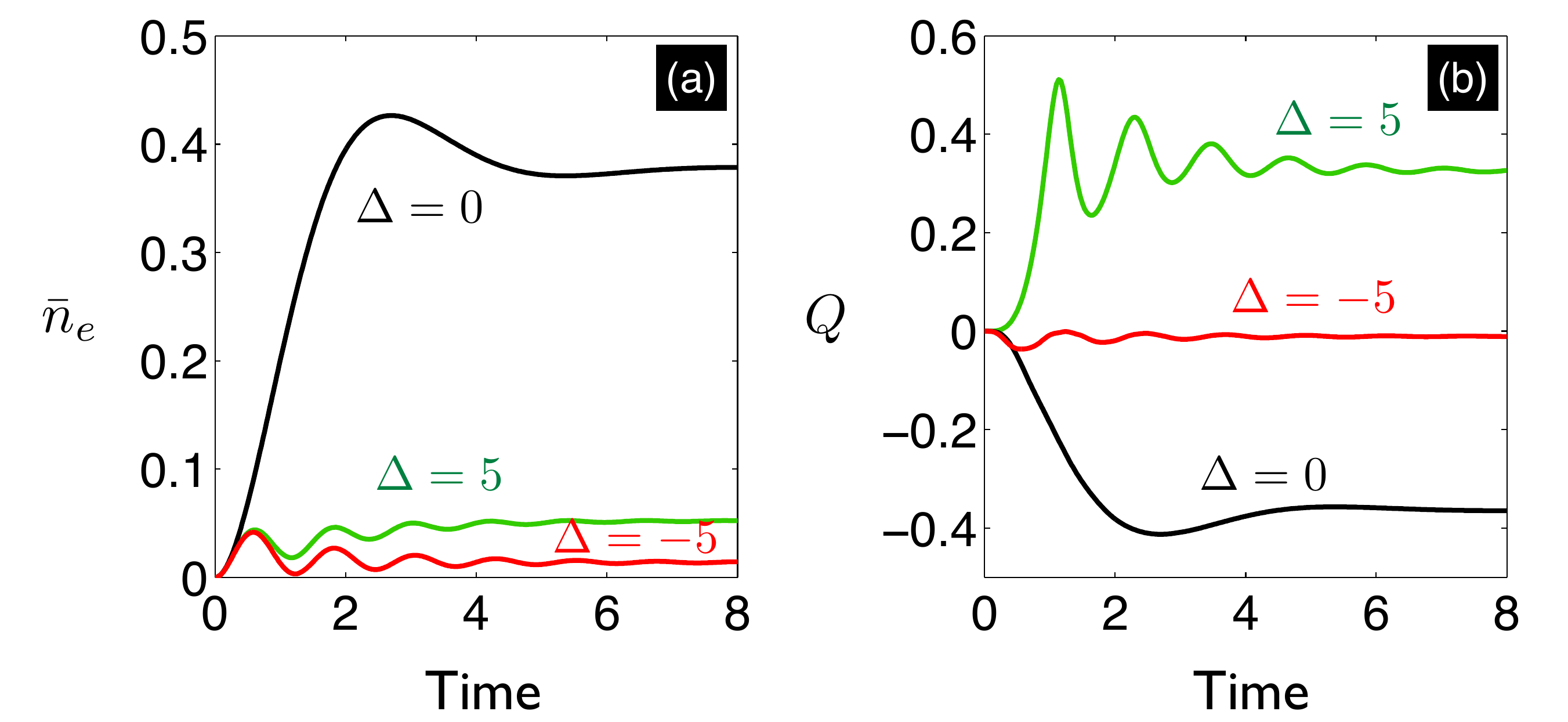} 

\caption{\label{fig1_Ne_Q} (Color online) (a) Mean Rydberg population $\bar{n}_{e}$ and (b)
Mandel $Q$ parameter as a function of time for laser detunings $\Delta=-5$,
$0$ and $5$ respectively. We fix units of $\gamma=1,$ taking $\lambda=0.6$,
$\chi=10$. In all cases the population of the double-excitation state
is less than $0.01$. For $\Delta=5$, the Rydberg excitation exhibits
super-Poissonian counting statistics with $Q>0$, while for $\Delta=-5$
and $\Delta=0$, the sub-Poissonian processes characterized by $Q<0$
are found. }
\end{figure}

We now introduce the collective spin operators $S^{+}=\sum_{j=1}^{N}|e_{j}\rangle\langle g_{j}|$
and $S^{-}=\sum_{j=1}^{N}|g_{j}\rangle\langle e_{j}|$, which according
to the Holstein-Primakoff transformation can be expressed in terms
of the bosonic operators $b$ and $b^{\dagger}$ (with $[b,b^{\dagger}]=1$)
as $S^{+}=b^{\dagger}\sqrt{N-b^{\dagger}b}$ and $S^{-}=\sqrt{N-b^{\dagger}b}b$
if the lowest energy level of these new operators is set to be the
atomic state in which all of the atoms are in the ground state {\cite{cpsun_prl2003,RevModPhys.82.1041}}.
It follows that $\sum_{j=1}^{N}|e_{j}\rangle\langle e_{j}|=\frac{1}{2}[S^{+},S^{-}]+\frac{N}{2}=b^{\dagger}b$
and $\sum_{j<k}|e_{j}\rangle\langle e_{j}|\otimes|e_{k}\rangle\langle e_{k}|=\frac{1}{2}b^{\dagger}b^{\dagger}bb$.
We then focus on the parameter regime
where the mean number of Rydberg excitations is much less than the
total number of atoms (i.e. $\bar{n}_{e}\equiv\langle b^{\dagger}b\rangle\ll N$)
resulting in $S^{+}\simeq\sqrt{N}b^{\dagger}$, $S^{-}\simeq\sqrt{N}b$
\cite{Schempp2014,Malossi2014}. Thus, the Hamiltonian of the system
can be rewritten by 
\begin{eqnarray}
H_{b} & = & -\Delta b^{\dagger}b+\frac{\chi}{2}b^{\dagger}b^{\dagger}bb+\lambda(b+b^{\dagger})\label{eq:H_boson}
\end{eqnarray}
with $\lambda=\sqrt{N}\Omega/2$. Note that the
system involving the Rydberg-Rydberg coupling behaves in resemblance
to the optical Kerr nonlinearity and its coherent dynamics will remain
in the symmetric Dicke state space. However, the spontaneous decay
from the Rydberg excited state (with the relaxation rate $\gamma$)
may lead to incoherent mixture with the asymmetric dark states. For
clarity, our work will focus on the state space mainly spanned by
null and single Rydberg excitation accompanied by a tiny fraction
of double excitations. The Rydberg population is confirmed by direct
simulations of the master equation $\dot{\rho}=-i[H_{b},\rho]+\frac{\gamma}{2}(2b\rho b^{\dagger}-\rho b^{\dagger}b-b^{\dagger}b\rho)$
for a zero-temperature thermal reservoir, and the counting statistics
of Rydberg excitation is quantified by the Mandel $Q$ parameter defined
as $Q=\langle(\Delta\hat{n}_{e})^{2}\rangle/\langle\hat{n}_{e}\rangle-1$
with $\hat{n}_{e}\equiv b^{\dagger}b$, as shown in Fig. \ref{fig1_Ne_Q}.
The exact simulation of the dissipative dynamics for few atoms with
the Hamiltonian Eq. (\ref{eq:H_MeanF}) can be found in Appendix \ref{apd_A},
which shows good agreement with the bosonization model. But
it should also be mentioned that for a realistic Rydberg system, such
as an atomic ensemble or a spin lattice, the finite interaction range
and the continuum of interatomic coupling strengths due to broad distribution
of atom position may induce loss of interatomic correlations, which
may negate the effectiveness of the model.

\section{Classical dynamics}

\begin{figure}
\includegraphics[width=1\columnwidth]{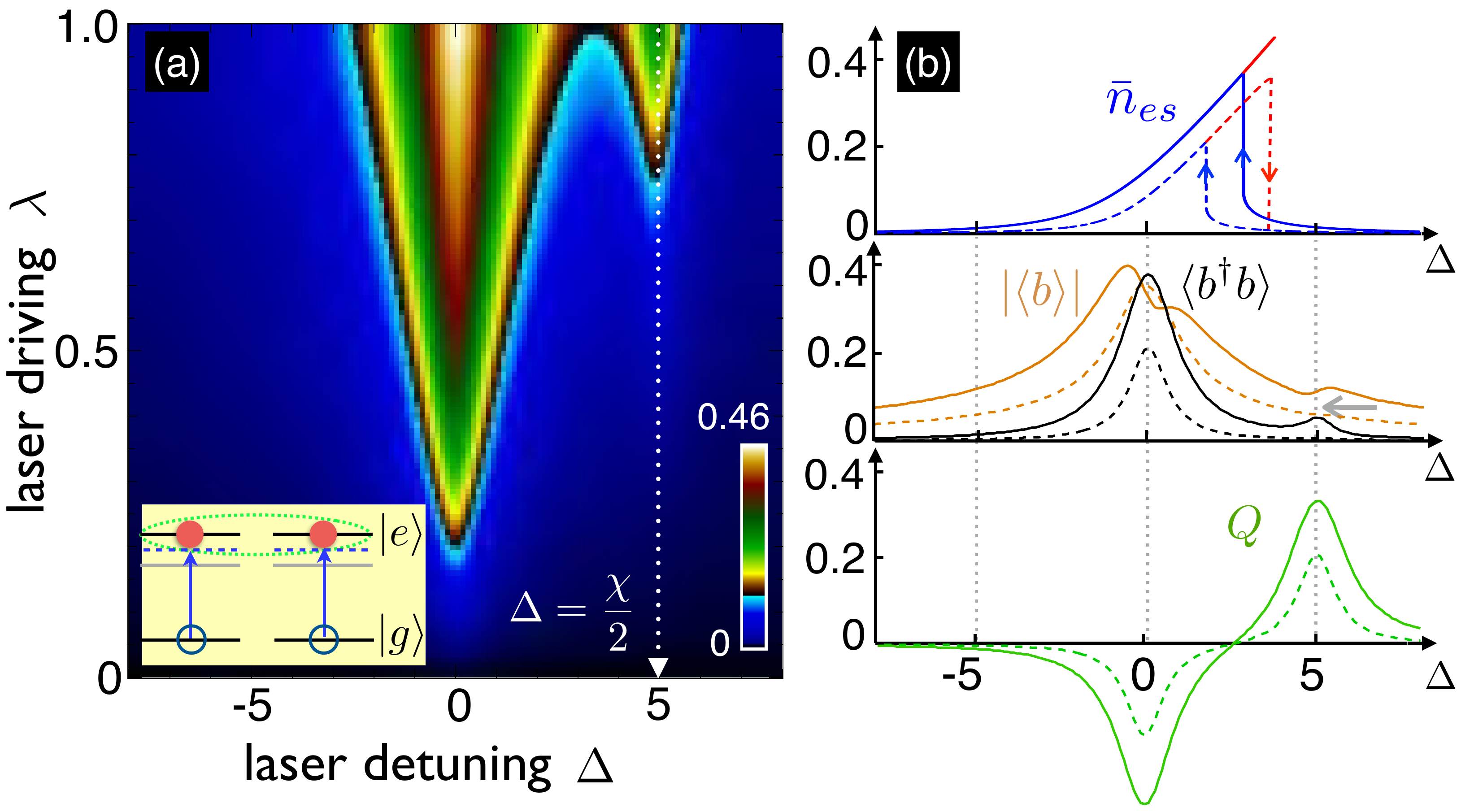}

\caption{\label{fig: Phasediagram}(Color online) (a) Mean number of Rydberg excitations $\langle b^{\dagger}b\rangle$
in the steady state with full quantum theory as a function of $\lambda$
and $\Delta$. The inset shows the pair excitation of two atoms for
$\Delta=\chi/2$. (b) Classical stationary mean $\bar{n}_{es}$, quantum
mechanical steady-state solution of $\langle b^{\dagger}b\rangle$
and $|\langle b\rangle|$, and Mandel $Q$ parameter versus laser
detuning $\Delta$ with $\lambda=0.3$ (dash) and $\lambda=0.6$ (solid).
Further parameters as in Fig. \ref{fig1_Ne_Q}. $\chi$, $\lambda$ and $\Delta$ are in units of $\gamma$.}
\end{figure}

The quantum Langevin equation of motion for the bosonic system is
\begin{eqnarray}
\dot{b} & = & -i\lambda-i\Gamma b-i\chi b^{\dagger}b^{2}+\xi(t)\label{eq:QLangevin}
\end{eqnarray}
where $\Gamma=-(\Delta+i\gamma/2)$ might include the decoherence
due to the laser linewidths and doppler broadening, and $\xi(t)$
is the zero-value-mean noise operator. Beyond the mean-field theory,
here the correlation between atoms is retained in the evolutional
dynamics. In the classical limit, the equation of motion for $\bar{\alpha}=\langle b\rangle$
($\bar{\alpha}^{*}=\langle b^{\dagger}\rangle$) are given by
\[
\dot{\bar{\alpha}}=-i\lambda-i\bar{\alpha}(\Gamma+\chi|\bar{\alpha}|^{2}),
\]
\begin{equation}
\dot{\bar{\alpha}}^{*}=i\lambda+i\bar{\alpha}^{*}(\Gamma^{*}+\chi|\bar{\alpha}|^{2}).\label{eq:CLangevin}
\end{equation}
Then, the mean number of Rydberg excitations $\bar{n}_{es}=|\bar{\alpha}_{s}|^{2}$
in the steady state fulfills the algebraic equation 
\begin{equation}
\chi^{2}\bar{n}_{es}^{3}-2\chi\Delta\bar{n}_{es}^{2}+\bar{n}_{es}|\Gamma|^{2}-\lambda^{2}=0\label{eq:Ne_ss}
\end{equation}
that allows at most three real positive roots. A stable $\bar{n}_{es}$
must conform to the Hurwitz criterion (given by $\partial(\lambda^{2})/\partial\bar{n}_{es}=3\chi^{2}\bar{n}_{es}^{2}-4\bar{n}_{es}\chi\Delta+|\Gamma|^{2}>0$
here), which ensures that the system returns to the stable branch
soon after the small perturbation. Therefore, if the conditions $|\Delta|>\sqrt{3}\gamma/2$
and $\Delta\chi>0$ are satisfied, the bistable region will be given
by $\bar{n}_{es}>\bar{n}_{es}^{(+)}$ or $\bar{n}_{es}<\bar{n}_{es}^{(-)}$
with the real positive $\bar{n}_{es}^{(\pm)}=(2\Delta\pm\sqrt{\Delta^{2}-\frac{3}{4}\gamma^{2}})/3\chi$.
For our interest, the dependence of $\bar{n}_{es}$ on the laser detuning
$\Delta$ for a repulsive interatomic interaction ($\chi>0$) is shown
in Fig. \ref{fig: Phasediagram}(b), which exhibits optical bistability
with hysteresis \cite{PhysRevLett.111.113901}. \textcolor{black}{The
coexistence of a low and a high Rydberg population in the off-resonance
regime ($\Delta>0$) leads to the bimodal counting distributions of
the Rydberg excitation \cite{Malossi2014}.} While for
$\Delta=0$, the bimodality can not arise due to the impossible bistability.

\section{Linearized dynamics}

The dissipative quantum dynamics for the system can be described by
the master equation or the corresponding Fokker-Planck equation (FPE)
\cite{Drummond_JPA} 
\begin{eqnarray}
\dot{P}(\alpha,\beta,t) & = & i[\frac{\partial}{\partial\alpha}(\Gamma\alpha+\chi\alpha^{2}\beta+\lambda)-\frac{\chi}{2}\frac{\partial^{2}}{\partial\alpha^{2}}\alpha^{2}\nonumber \\
 &  & -\frac{\partial}{\partial\beta}(\Gamma^{*}\beta+\chi\beta^{2}\alpha+\lambda)+\frac{\chi}{2}\frac{\partial^{2}}{\partial\beta^{2}}\beta^{2}]P(\alpha,\beta,t)\nonumber \\
\label{eq:FPE}
\end{eqnarray}
by introducing the non-diagonal generalized $P$ representation defined
by 
\begin{equation}
\rho(t)=\int_{D}d\sigma(\alpha,\beta)P(\alpha,\beta,t)\frac{|\alpha\rangle\langle\beta^{*}|}{\langle\beta^{*}|\alpha\rangle},
\end{equation}
where $D$ is the integration domain, and $d\sigma$ is the integration
measure that can be a volume integral $d^{2}\alpha d^{2}\beta$ over
a complex phase space or a line integral $d\alpha d\beta$ over a
manifold embedded in a complex phase space \cite{Drummond_JPA}. Note
that $\alpha$, $\beta$ are arguments of the generalized $P$ function
(in correspondence to the $c$ numbers of $b$, $b^{\dagger}$) and
are complex conjugate in the mean $\bar{\alpha}=\bar{\beta}^{*}$.

Transforming the FPE into the Ito form yields a pair of stochastic
differential equations (SDEs):
\begin{eqnarray}
\frac{\partial}{\partial t}\left[\begin{array}{c}
\alpha\\
\beta
\end{array}\right] & = & \left[\begin{array}{c}
-i(\Gamma\alpha+\chi\alpha^{2}\beta+\lambda)\\
i(\Gamma^{*}\beta+\chi\beta^{2}\alpha+\lambda)
\end{array}\right]\nonumber \\
 &  & +\left[\begin{array}{cc}
-i\chi\alpha^{2} & 0\\
0 & i\chi\beta^{2}
\end{array}\right]^{\frac{1}{2}}\left[\begin{array}{c}
\xi_{1}(t)\\
\xi_{2}(t)
\end{array}\right]\label{eq:Ito_SWF}
\end{eqnarray}
with $\xi_{1,2}(t)$ the delta-correlated Gaussian white noise, satisfying
$\langle\xi_{1,2}(t)\rangle=0$ and $\langle\xi_{1,2}(t)\xi_{1,2}(t^{\prime})\rangle\sim\delta(t-t^{\prime})$.
We then linearize the SDEs around the classically steady-state solution
$\bar{\alpha}_{s}(\bar{\beta}_{s}=\bar{\alpha}_{s}^{*})$, and find
that the quantum fluctuation $\delta\mathbf{\hat{X}}=[\delta\hat{\alpha},\delta\hat{\beta}]^{T}$
close to the classical stable branches obeys the equation 
\begin{equation}
\delta\dot{\mathbf{\hat{X}}}=-\mathbf{\boldsymbol{\mu}}(\bar{\alpha}_{s})\cdot\mathbf{\delta\mathbf{\hat{X}}}+\mathbf{D}^{1/2}(\bar{\alpha}_{s})\cdot\boldsymbol{\xi}(t),\label{eq:SDE_linearize}
\end{equation}
where 
\begin{equation}
\boldsymbol{\mathbf{\mu}}(\bar{\alpha}_{s})=-i\left[\begin{array}{cc}
\Gamma+2\chi|\bar{\alpha}_{s}|^{2} & -\chi\bar{\alpha}_{s}^{*2}\\
\chi\bar{\alpha}_{s}^{2} & -\Gamma-2\chi|\bar{\alpha}_{s}|^{2}
\end{array}\right]
\end{equation}
 and 
\begin{equation}
\mathbf{D}(\bar{\alpha}_{s})=\left[\begin{array}{cc}
-i\chi\bar{\alpha}_{s}^{2} & 0\\
0 & i\chi\bar{\alpha}_{s}^{*2}
\end{array}\right]
\end{equation}
are the linearized drift and the diffusion array, respectively. The
correlation matrix with regard to $\delta\mathbf{\hat{X}}$ can now
be calculated through
\begin{eqnarray}
\mathbf{C} & = & \left[\begin{array}{cc}
\langle b^{2}\rangle-\langle b\rangle^{2} & \langle b^{\dagger}b\rangle-|\langle b\rangle|^{2}\\
\langle b^{\dagger}b\rangle-|\langle b\rangle|^{2} & \langle b^{\dagger2}\rangle-\langle b^{\dagger}\rangle^{2}
\end{array}\right]\nonumber \\
 & = & \frac{\mathbf{D}\cdot\textrm{Det}(\mathbf{\boldsymbol{\mu}})+[\boldsymbol{\mu-}\mathbf{I}\cdot\textrm{Tr}(\boldsymbol{\mu})]\mathbf{D}[\boldsymbol{\mu-}\mathbf{I}\cdot\textrm{Tr}(\boldsymbol{\mu})]^{T}}{2\textrm{Tr}(\boldsymbol{\mu})\textrm{Det}(\boldsymbol{\mu})}.
\end{eqnarray}

The above result allows us to obtain the mean Rydberg population and
the Mandel $Q$ parameter in the linearized regime: (see Appendix \ref{apd_B})
\begin{eqnarray}
\bar{n}_{e} & \approx & |\bar{\alpha}_{s}|^{2}+\langle\delta\hat{\alpha}^{\dagger}\delta\hat{\alpha}\rangle=\bar{n}_{es}+\frac{\chi^{2}}{2\Lambda}\bar{n}_{es}^{2},\label{eq:ne_linearize}
\end{eqnarray}

\begin{eqnarray}
Q & \approx & 2\bar{n}_{es}\textrm{Re}[\langle\delta\hat{\alpha}^{2}\rangle/\bar{\alpha}_{s}^{2}]+2\langle\delta\hat{\alpha}^{\dagger}\delta\hat{\alpha}\rangle\nonumber \\
 & = & \frac{\bar{n}_{es}\chi}{\Lambda}(\Delta-\chi\bar{n}_{es}),\label{eq:Q_linearize}
\end{eqnarray}
with $\Lambda=|\Gamma|^{2}-4\chi\Delta\bar{n}_{es}+3\chi^{2}\bar{n}_{es}^{2}.$
The steady-state solution $\bar{n}_{es}$ can be obtained by solving
the set of classically nonlinear equations (\ref{eq:CLangevin}),
and there may exist two stable solution for a well-selected laser
detuning as shown in Fig.\ref{fig: Phasediagram}(b). We now assume
this solution has been found and focus on the effect of the quantum
fluctuations. One of the intriguing effects can be found in the asymptotic
expansion of $\bar{n}_{e}$ to the first-order {[}see Eq. (\ref{eq:ne_linearize}){]},
where the Rydberg population is enhanced by the Rydberg-Rydberg coupling
via the intensity of quantum fluctuation $\langle\delta\hat{\alpha}^{\dagger}\delta\hat{\alpha}\rangle$
\cite{PhysRevLett.108.023602,Garttner2014a}. Second, due to the Rydberg-Rydberg
interactions, the laser field drives the atomic transition from the
ground state to the Rydberg state in a cooperative manner, with the
fluctuations of the collective excitation number being characterized
by the Mandel $Q$ factor. The results show that the collective excitation
number exhibits super-Poissonian distribution for the laser detuning
$\Delta$ greater than the collective energy shift $\bar{n}_{es}\chi$
among the correlated interacting particles ($\Delta/\bar{n}_{es}\chi>1$)
and sub-Poissonian distribution for the other ($\Delta/\bar{n}_{es}\chi<1$),
leading to collective quantum jumps while the system stays in the
two classically stable branches \cite{PhysRevLett.108.023602}. The
linear theory breaks down for $\Lambda\rightarrow0$ (corresponding
to the violation of the Hurwitz criterion), in which case $\bar{n}_{es}$
approaches the onset of instability.

\section{Full quantum dynamics}

Having seen the effect of the quantum fluctuations based on the linearization,
we next address the question with full quantum theory. In the quantum
noise limit, there exists an exact steady-state solution $P_{ss}(\alpha,\beta)=\alpha^{c-2}\beta^{d-2}exp[-z(1/\alpha+1/\beta)+2\alpha\beta$
for the FPE (\ref{eq:FPE}) with $z=2\lambda/\chi,$ $c=2\Gamma/\chi$,
and $d=c^{*}$ (see Appendix \ref{apd_C}). While we obtain the generalized
$P$ function, the Rydberg population and the Rydberg counting statistics
can be calculated via 
\begin{equation}
\langle b^{\dagger n}b^{m}\rangle=\frac{\int_{D}d\sigma(\alpha,\beta)P_{ss}(\alpha,\beta)\beta^{n}\alpha^{m}}{\int_{D}d\sigma(\alpha,\beta)P_{ss}(\alpha,\beta)}.
\end{equation}
The appropriate integration domain $D$ here will be a complex manifold
embedded in the space $\mathbb{C}^{2}$ and each path of integration
is chosen to be a Hankel path $\mathcal{O}$. These ensure that the
distribution function $P(\alpha,\beta)$ vanishes correctly at the
boundary. Skipping over the fussy calculations we finally come at
(see Appendix \ref{apd_C})
\begin{equation}
\langle b\rangle=\frac{z}{c}\frac{_{0}F_{2}(c+1,d,2z^{2})}{_{0}F_{2}(c,d,2z^{2})}, \label{eq:b_fp}
\end{equation}

\begin{equation}
\langle b^{\dagger}b\rangle=\frac{z^{2}}{cd}\frac{_{0}F_{2}(c+1,d+1,2z^{2})}{_{0}F_{2}(c,d,2z^{2})}, \label{eq:ne_fp}
\end{equation}
\begin{eqnarray}
Q & = & \frac{z^{2}}{(c+1)(d+1)}\frac{_{0}F_{2}(c+2,d+2,2z^{2})}{_{0}F_{2}(c+1,d+1,2z^{2})}\nonumber \\
 &  & -\frac{z^{2}}{cd}\frac{_{0}F_{2}(c+1,d+1,2z^{2})}{_{0}F_{2}(c,d,2z^{2})}, \label{eq:Q_fp}
\end{eqnarray}
where $_{0}F_{2}(c,d,z)$ is a hypergeometric series defined as
\begin{equation}
_{0}F_{2}(c,d,z)=\sum_{n=0}^{\infty}\frac{z^{n}}{n!}\frac{\Gamma(c)\Gamma(d)}{\Gamma(c+n)\Gamma(d+n)}
\end{equation}
with $\Gamma(x)=[\frac{z^{1-x}}{2\pi i}\int_{\mathcal{O}}\sigma{}^{-x}e^{\sigma z}d\sigma]^{-1}$
the gamma-function. 

The phase diagram for the mean number of Rydberg excitations in ($\lambda,\Delta$)
space shows that the Rydberg excitation dramatically increases at
the laser detunings $\Delta=0$ and $\Delta=\chi/2$ {[}see Fig. \ref{fig: Phasediagram}(a){]},
which corresponds to individually resonant excitation of the each
atom and pair excitation of the interacting atoms, respectively. Numerical
simulation proves that the system is confined in the subspace spanned
by the null- and double-excitation states for $\Delta=\chi/2$ without
considering the relaxation, confirming that for this detuning the
atomic excitation is caused by the two-photon process; this is fundamentally
different from the case of resonant driving, where the system has
no probability of being pumped to the state with more than one excitation
due to the Rydberg blockade. Taking into account the atomic spontaneous
emission, the issue whether the many-body states of the Rydberg atoms
are created by the coherent two-photon process or sequential excitations
of individual atoms is then determined by the sequential to two-photon
ratio, which is proportional to $\gamma^{2}/N\Omega^{2}$ \cite{Schempp2014}.
Thus, the system favors double excitation since the multiatom coherence
becomes significant and the dephasing induced by random distribution
of atom position and specific spatial laser profile is neglected \cite{Schonleber2014b}.
This can be further confirmed by measuring the Mandel $Q$ parameter
that is in close relation to the spatial correlation function $g^{(2)}(R)$
\cite{Wuster2010a}. The excitation processes exhibit the sub-Poissonian
character ($Q<0$) on resonance and super-Poissonian character ($Q>0$)
for the other. Besides, due to the fact that \textcolor{black}{the
interatomic correlations between any pairs of atoms are included,
the cooperative transitions from the doubly excitation state to the
higher excitation states are largely detuned from resonance and are
suppressed for the collective energy shifts being much larger than
the laser driving strength.} 

For $0<\Delta<\chi/2$, both the individual
excitation and the pair excitation are possible. While the classical
dynamics predicts an optical bistability, the quantum mechanical calculation
including the interatomic correlation does not exhibit bistability.
It should also be noted that the pair excitation of two atoms is a
nonclassical effect that is unable to be revealed by the classical
theory {[}see arrow in Fig. \ref{fig: Phasediagram}(b){]} (see Appendix \ref{apd_D}).
An extra quantum signature for the system is given by the dips in
the laser-detuning dependence of $|\langle b\rangle|$, which arise
at the transition points of Mandel $Q$ factor that shows a reduced
or enhanced quantum fluctuations of Rydberg excitations \cite{PhysRevA.84.031402,Ates2012,PhysRevLett.108.023602}.

\begin{figure}
\begin{centering}
\includegraphics[width=0.8\columnwidth]{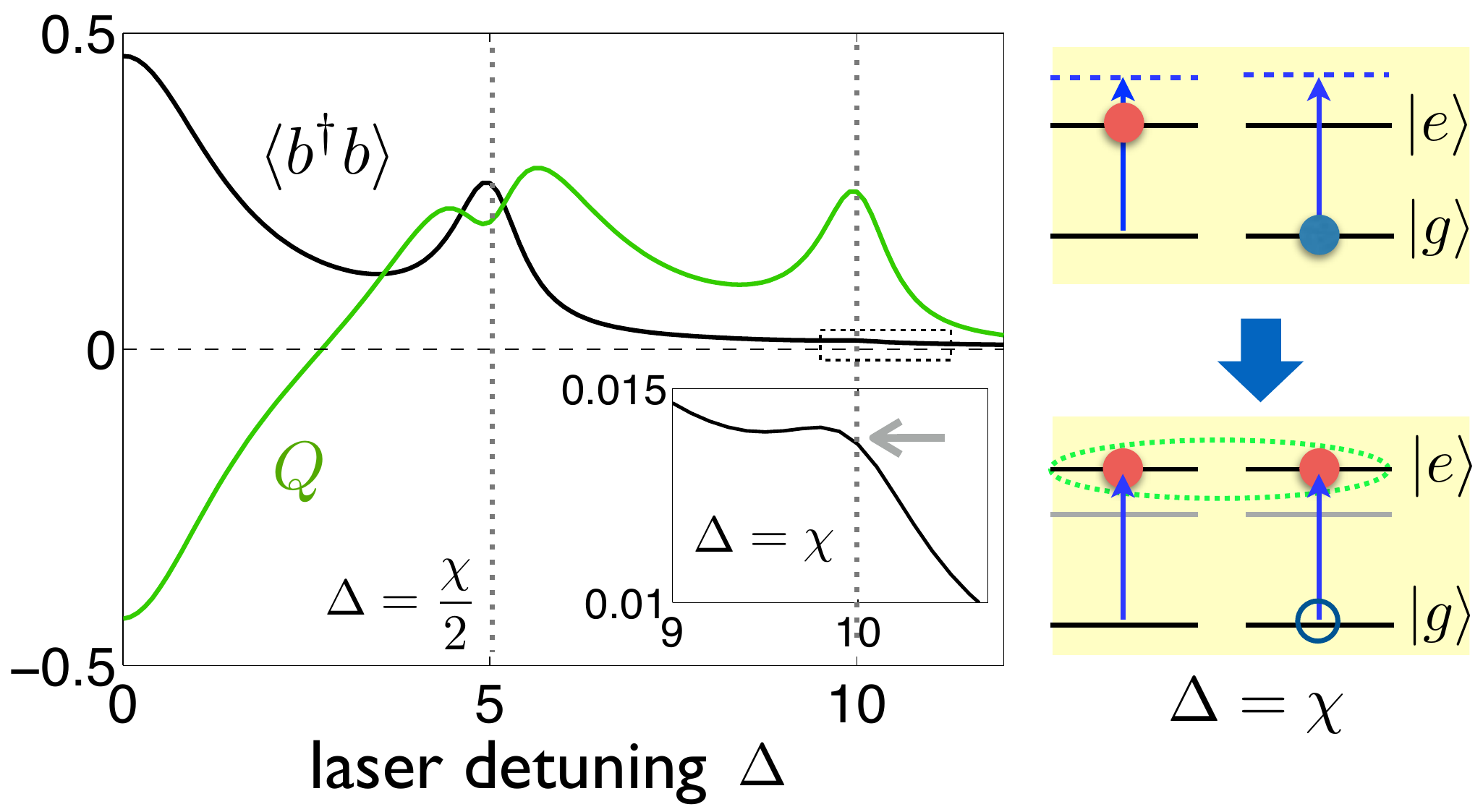}
\par\end{centering}
\caption{\label{fig:ne=000026Q_L}(Color online) Left panel: Quantum mechanical steady-state
solution of mean Rydberg population $\langle b^{\dagger}b\rangle$
and Mandel $Q$ parameter versus laser detuning $\Delta$ with increasing
driving strength $\lambda=1$ and $\chi=10$. $\chi$, $\lambda$ and $\Delta$ are in units of $\gamma$. Right panel: An already
excited Rydberg atom forwards the excitation of the surrounding atoms
interacting with it for $\Delta=\chi$. See the text for detail.}
\end{figure}

Increasing the laser intensity will enhance the mean number of Rydberg
excitations for $\Delta=\chi/2$, but meanwhile weaken the super-Poissonian
excitation process due to the decreasing ratio $\Delta/\bar{n}_{es}\chi$,
as shown in Fig. \ref{fig:ne=000026Q_L}. This fact can be directly
understood from the linearized calculation Eq. (\ref{eq:Q_linearize}){]}.
For the far off-resonance regime $\Delta\gg\Omega$, there is a small
chance (roughly $N\Omega^{2}/4\Delta^{2}$) for an atom being excited
to the Rydberg state if all the atoms are initially in the ground
state. However, while an atom has been luckily excited by the laser
beam with the detuning $\Delta=\chi$, then similar to the Rydberg
aggregates, the surrounding atoms tend to be excited cooperatively
due to the Rydberg-Rydberg coupling, as sketched in the right panel
of Fig. \ref{fig:ne=000026Q_L}. Besides, the sequential to simultaneous
three-photon excitation ratio is on the order of $\gamma^{2}\Delta^{2}/N^{2}\Omega^{4}$
and is extremely weak here. 

\section{ Discussion and Conclusion}

As mentioned in Sec. II, the simplified picture here requires the interatomic correlations to be well preserved amongst the whole Rydberg ensemble, and the only scenario where the present model describes actual Rydberg atom interactions is when all interactions are so strong ($\chi\gg\Delta$ or $\lambda$) that the ensemble is fully blockaded, i.e. not even double excitations can ever occur. While the scale of a Rydberg atom system is much larger than the blockade radius (i.e. the critical correlation distance), the quantum correlations may vanish for Rydberg atoms seated far away from each other; then the situation becomes different for the Rydberg excitation number being related to the number of blockade spheres that fit into the excitation volume. The excitation dynamics for ultracold Rydberg gas with high atomic density has been experimentally studied in the strong dephasing regime (due to the spatial dependence of atom positions or laser geometry) and in the limit of short excitation times \cite{Schempp2014,Malossi2014}.

It is more interesting to study the excitation dynamics in the scenario that $\chi$ is of a comparable order to $\Delta$ or $\lambda$ as employed here. The quantum correlations among all the atoms become essential, giving rise to the pronounced quantum signature discussed previously. But for actual experiments such as individual Rydberg atoms trapped in tunable two-dimensional arrays of optical microtraps \cite{Labuhn_Nature2015}, the distance-dependent interatomic coupling strength is determined by the specific lattice sites that the atoms locate on, and therefore the factor of 8 (or higher) difference in interactions comparing two adjacent atoms already introduces correlations and breaks the present approach. This is regardless of the dimensionality or geometry of a lattice. However, the strong two-photon transitions will be still visible while many pairs of the atoms feature distance that fits in with the two-photon resonance condition, namely, the two-photon resonance can be enhanced by choosing the appropriate detuning for a given lattice constant \cite{Schonleber2014b}. The limitations arise from the fact that the atoms near the edge of lattice evolve as an inhomogeneous distribution of atom position, which breaks the spatial symmetry and may induce an additional dephasing reducing the visibility of two-photon transitions \cite{Labuhn_Nature2015}. 

In conclusion, we have analyzed the steady-state excitation dynamics of a weakly-driven Rydberg-atom system by treating the inhomogeneous Rydberg-Rydberg interaction as an all-to-all coupling. While the present model can not represent the full picture of a realistic system, in contrast to the previous analytical or experimental studies that mainly focus on the mean-field dynamics and neglect the interatomic correlations, the analytical calculation here allows us to find the effect of the quantum fluctuations on the Rydberg excitation process and the quantum phenomena such as simultaneous bi-excitation of Rydberg atoms and interaction-assisted Rydberg excitation, which are referred to as the quantum signature of the system. The simplified model forms the basis for quantitatively understanding the Rydberg aggregates \cite{Schempp2014,Urvoy2014a}, bimodal counting distribution \cite{Malossi2014}, and kinetic constraints \cite{Valado2015} in recent experimental observations with Rydberg atoms, and more generally the many-body effect for the interacting spin-$1/2$ systems \cite{Britton_Nature2012}. 

\section*{Acknowledgments}

We acknowledge W. Li for helpful discussions. This work
was supported by the Major State BasicResearch Development
Program of China under Grant No. 2012CB921601; the
National Natural Science Foundation of China under Grants
No. 11305037, No. 11374054, No. 11534002, No. 11674060,
No. 11575045, and No. 11405031; and the Natural Science
Foundation of Fujian Province under Grant No. 2013J01012.

\appendix

\section{\label{apd_A}The exact simulation of the dissipative dynamics with
atomic spin operators}

In this section, we show that the exact simulation of the Hamiltonian
$H$ in the main text with spin operators for few atoms is found to
be in good agreement with the theoretical model in terms of bosonic
operators. We also show how to obtain the correlation matrix under
linearization and the analytical steady-state solution for the Fokker-Planck
equation as in ref. \cite{Drummond_JPA} discussing the optical bistability
for nonlinear polarisability model. The nonclassical nature of the
quantum mechanical steady states is also shown by the Wigner functions.

\begin{figure}
\begin{centering}
\includegraphics[width=1\columnwidth]{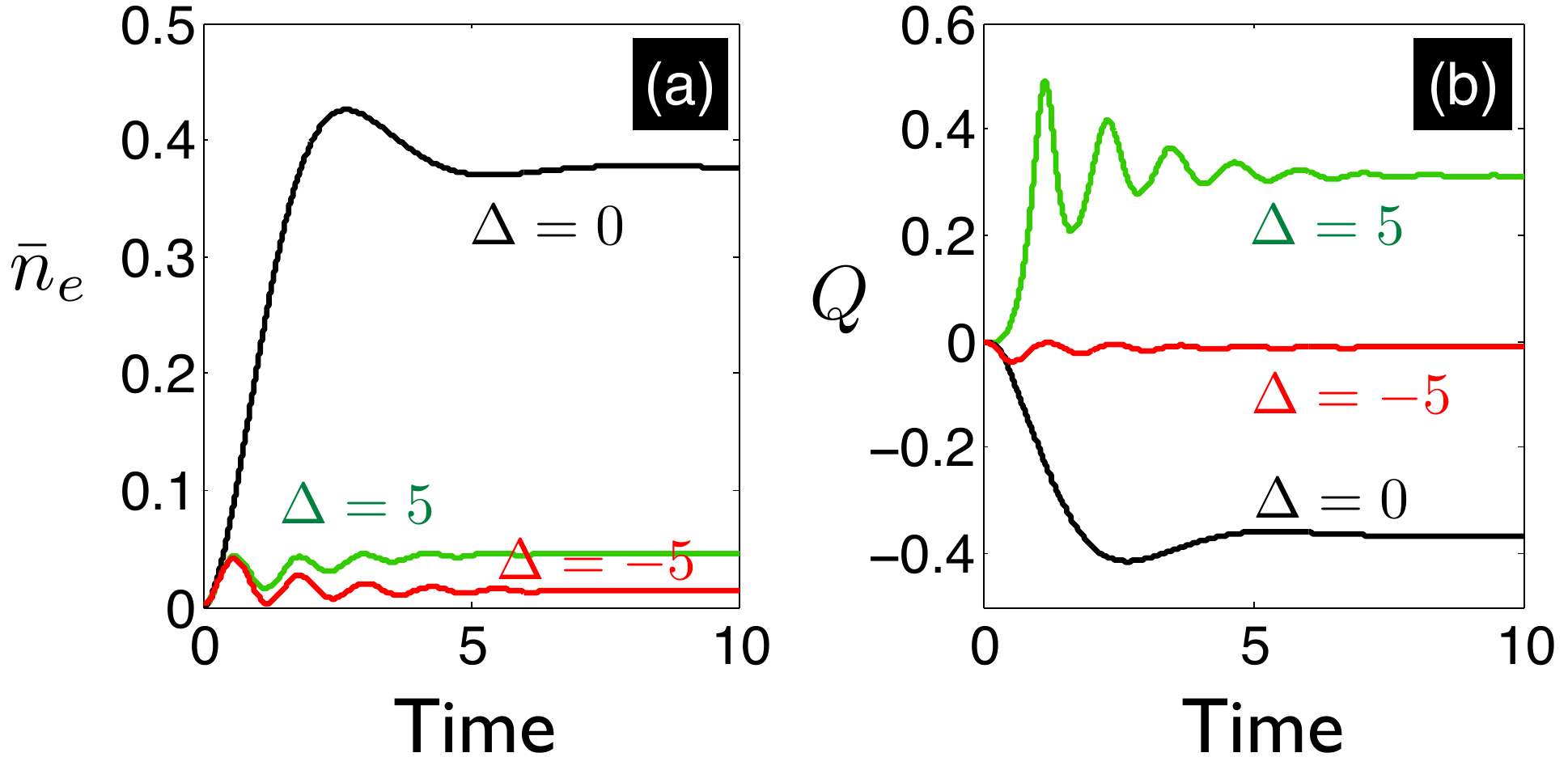}
\par\end{centering}

\centering{}\caption{\label{fig1_s}(Color online) (a) Mean Rydberg population $\bar{n}_{e}$ and (b)
Mandel $Q$ parameter as a function of time for laser detunings $\Delta=-5$,
$0$ and $5$ respectively. Further parameters as in Fig. 1 in the
main text.}
\end{figure}

The exactly dissipative dynamics of the laser-driven system with the
Hamiltonian $H$ in the main text can be described by the master equation
\begin{equation}
\dot{\rho}=\frac{1}{i\hbar}[H,\rho]+\frac{\gamma}{2}\sum_{j=1}^{N}\mathcal{L}_{j},\label{eq:master_s}
\end{equation}
where
\begin{equation}
\mathcal{L}_{j}=2\sigma_{ge}^{(j)}\rho\sigma_{eg}^{(j)}-\sigma_{eg}^{(j)}\sigma_{ge}^{(j)}\rho-\rho\sigma_{eg}^{(j)}\sigma_{ge}^{(j)},
\end{equation}
$ $$\sigma_{ge}^{(j)}=|g_{j}\rangle\langle e_{j}|$ and $ $$\sigma_{eg}^{(j)}=|e_{j}\rangle\langle g_{j}|$.
With the excitation-number operator being $\hat{n}_{e}=\sum_{j=1}^{N}|e_{j}\rangle\langle e_{j}|$,
the mean number of Rydberg excitations and the Mandel $Q$ parameter
can be calculated via $\bar{n}_{e}=\textrm{Tr}(\rho\hat{n}_{e})$
and $Q=\textrm{Tr}(\rho(\Delta\hat{n}_{e})^{2})/\textrm{Tr}(\rho\hat{n}_{e})-1$,
respectively. 

\begin{figure}
\begin{centering}
\includegraphics[width=1\columnwidth]{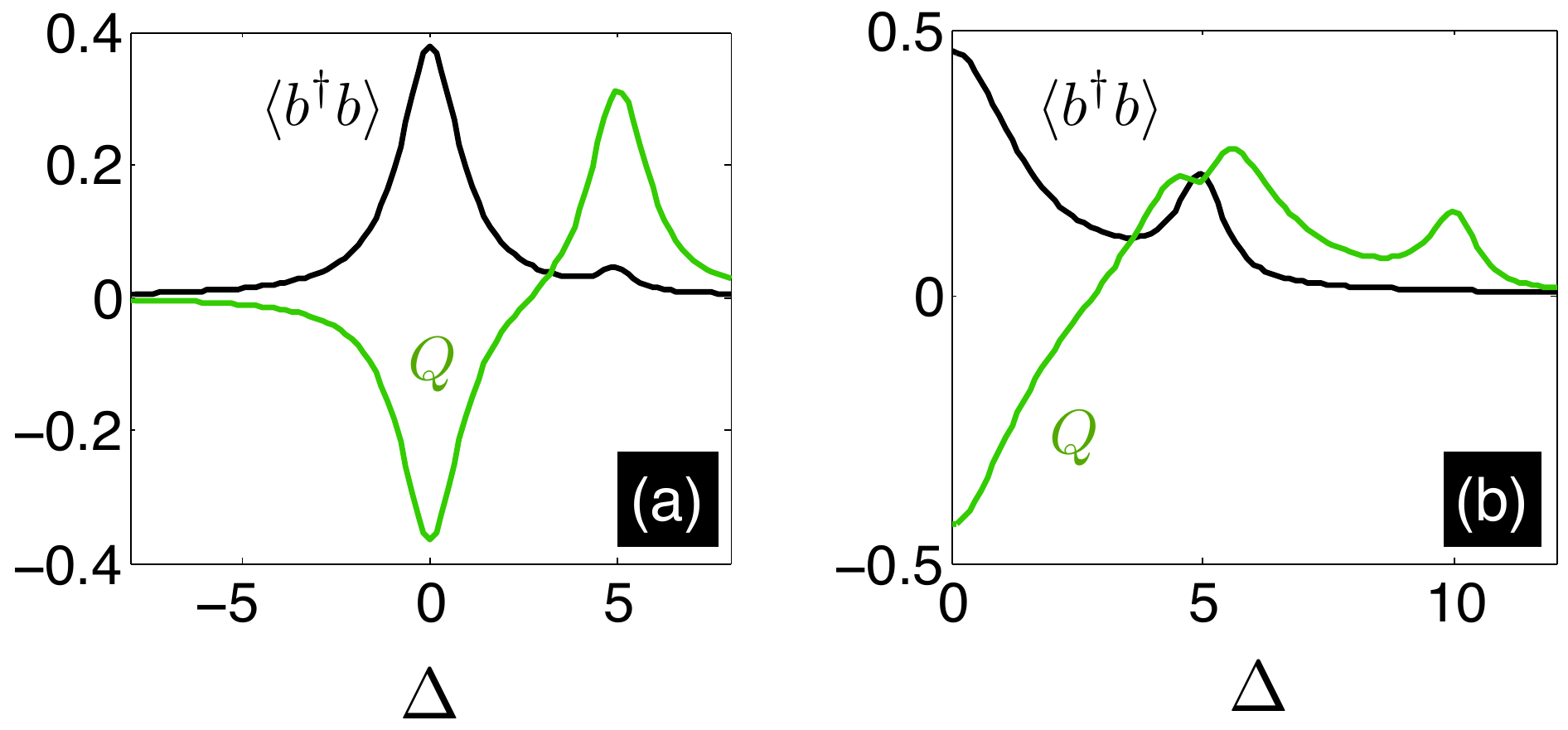}
\par\end{centering}

\caption{\label{fig2_s}(Color online) Quantum mechanical steady-state solution of $\bar{n}_{e}$
and Mandel $Q$ parameter versus laser detuning $\Delta$ for (a)
$\lambda=0.6$ and (b) $\lambda=1$. Further parameters as in Fig.
2(b) in the main text.}
\end{figure}

Simulating the Eq. (\ref{eq:master_s}) with $N=6$, we show the time-dependent
mean Rydberg population $\bar{n}_{e}$ and Mandel $Q$ parameter in
Fig. \ref{fig1_s}, and the steady-state solution of $\bar{n}_{e}$
and $Q$ versus laser detuning $\Delta$ in Fig. \ref{fig2_s}, which
demonstrate the quantitatively good agreement with the results obtained
via the simulation of the bosonization model.

\section{\label{apd_B}The quantum fluctuation under linearization}

Splitting the operators $b$ ($b^{\dagger}$) into classical and quantum
parts $\alpha=\bar{\alpha}+\delta\hat{\alpha}(t)$, $\beta=\bar{\beta}+\delta\hat{\beta}(t)$
($\bar{\alpha}^{*}=\bar{\beta}$), we obtain the linearized correlation
matrix given by 
\begin{eqnarray}
\mathbf{C} & = & \left[\begin{array}{cc}
\langle b^{2}\rangle-\langle b\rangle^{2} & \langle b^{\dagger}b\rangle-|\langle b\rangle|^{2}\\
\langle b^{\dagger}b\rangle-|\langle b\rangle|^{2} & \langle b^{\dagger2}\rangle-\langle b^{\dagger}\rangle^{2}
\end{array}\right]\nonumber \\
 & = & \left[\begin{array}{cc}
\langle\delta\hat{\alpha}^{2}\rangle & \langle\delta\hat{\alpha}^{\dagger}\delta\hat{\alpha}\rangle\\
\langle\delta\hat{\alpha}^{\dagger}\delta\hat{\alpha}\rangle & \langle\delta\hat{\alpha}^{\dagger2}\rangle
\end{array}\right]
\end{eqnarray}
where
\begin{equation}
\langle\delta\hat{\alpha}^{2}\rangle=-\frac{\chi\bar{\alpha}^{2}(\Gamma^{*}+2\chi\bar{n}_{es})}{2\Lambda},
\end{equation}
\begin{equation}
\langle\delta\hat{\alpha}^{\dagger}\delta\hat{\alpha}\rangle=\frac{\chi^{2}\bar{n}_{es}^{2}}{2\Lambda}
\end{equation}
and
\begin{equation}
\langle\delta\hat{\alpha}^{\dagger2}\rangle=-\frac{\chi\bar{\alpha}^{*2}(\Gamma+2\chi\bar{n}_{es})}{2\Lambda}
\end{equation}
with $\Lambda=\frac{\gamma^{2}}{4}+\Delta^{2}-4\chi\Delta\bar{n}_{es}+3\chi^{2}\bar{n}_{es}^{2}$
are obtained from $\mathbf{C}=\frac{\mathbf{D}\cdot\textrm{Det}(\mathbf{\boldsymbol{\mu}})+[\boldsymbol{\mu-}\mathbf{I}\cdot\textrm{Tr}(\boldsymbol{\mu})]\mathbf{D}[\boldsymbol{\mu-}\mathbf{I}\cdot\textrm{Tr}(\boldsymbol{\mu})]^{T}}{2\textrm{Tr}(\boldsymbol{\mu})\textrm{Det}(\boldsymbol{\mu})}$
by using the drift and diffusion arrays. Thus, the mean number of
Rydberg excitations and the Mandel $Q$ parameter, in the linearized
regime, are given by  
\begin{eqnarray}
\bar{n}_{e} & \approx & |\bar{\alpha}|^{2}+\langle\delta\hat{\alpha}^{\dagger}\delta\hat{\alpha}\rangle=\bar{n}_{es}+\frac{\chi^{2}}{2\Lambda}\bar{n}_{es}^{2}
\end{eqnarray}
and

\begin{eqnarray}
Q & = & \frac{[\langle(b^{\dagger}b)^{2}\rangle-\langle b^{\dagger}b\rangle^{2}]}{\langle b^{\dagger}b\rangle}-1\nonumber \\
 & = & \frac{2\textrm{Re}[\bar{\alpha}^{*2}\langle\delta\hat{\alpha}^{2}\rangle]+2\bar{n}_{es}\langle\delta\hat{\alpha}^{\dagger}\delta\hat{\alpha}\rangle}{\bar{n}_{es}+\langle\delta\hat{\alpha}^{\dagger}\delta\hat{\alpha}\rangle}+O[(\delta\hat{\alpha}^{\dagger})^{2}(\delta\hat{\alpha})^{2}],\nonumber \\
\end{eqnarray}
respectively. For $\langle\delta\hat{\alpha}^{\dagger}\delta\hat{\alpha}\rangle\ll\bar{n}_{es}$,
and keeping only the first-order fluctuation, the Mandel $Q$ parameter
approximates to 
\begin{eqnarray}
Q & \approx & 2\textrm{Re}[\langle\delta\hat{\alpha}^{2}\rangle/\bar{\alpha}^{2}]+2\langle\delta\hat{\alpha}^{\dagger}\delta\hat{\alpha}\rangle\nonumber \\
 & = & \frac{\bar{n}_{es}\chi}{\Lambda}(\Delta-\chi\bar{n}_{es}).
\end{eqnarray}
It should be realized that the mean Rydberg population is enhanced
by the intensity of quantum fluctuation $\langle\delta\hat{\alpha}^{\dagger}\delta\hat{\alpha}\rangle$
induced by the Rydberg-Rydberg interaction. Moreover, the Mandel $Q$
parameter is additionally affected by the quantum fluctuation $\langle\delta\hat{\alpha}^{2}\rangle$
around $\langle b\rangle$. The cooperation of the quantum fluctuations
$\langle\delta\hat{\alpha}^{2}\rangle$ and $\langle\delta\hat{\alpha}^{\dagger}\delta\hat{\alpha}\rangle$
finally determines the statistical dynamics of the Rydberg excitation,
in correspondence to the super(sub)-Poissonian counting statistics.
On the other hand, the Hurwitz criteria for stability is given by
$\Lambda>0$. While $\Lambda$ approaches to zero, the quantum fluctuations
diverges and the linearization fails.

\section{\label{apd_C}The exact solution of the Fokker-Planck equation}

Consider the Fokker-Planck equation (in the quantum noise limit, excluding
the thermal noise, i.e. $n_{th}=0$)
\begin{eqnarray}
\dot{P}(\alpha,\beta,t) & = & [\partial_{u}\mathbf{A}_{u}(\alpha,\beta)+\frac{1}{2}\partial_{u}\partial_{v}\mathbf{D}_{uv}(\alpha,\beta)]P(\alpha,\beta,t)\nonumber \\
 & = & \partial_{u}\{\frac{1}{2}\mathbf{D}_{uv}[(\mathbf{D}_{vu})^{-1}(2\mathbf{A}_{u}+\partial_{v}\mathbf{D}_{uv})P+\partial_{v}P\}\nonumber \\
\end{eqnarray}
with $u,v=\alpha,\beta$, and the diffusion array is
\begin{equation}
\mathbf{D}(\alpha,\beta)=\left[\begin{array}{cc}
-i\chi\alpha^{2} & 0\\
0 & i\chi\beta^{2}
\end{array}\right],
\end{equation}
\begin{equation}
\mathbf{A}(\alpha,\beta)=\left[\begin{array}{c}
i(\Gamma\alpha+\chi\alpha^{2}\beta+\lambda)\\
-i(\Gamma^{*}\beta+\chi\beta^{2}\alpha+\lambda)
\end{array}\right].
\end{equation}

A steady-state exact solution for the Fokker-Planck equation exists
only when the conditional equations $\partial_{u}V_{v}=\partial_{v}V_{u}$
are fulfilled \cite{RevModPhys.47.67}, where the potential function
is given by $V_{\varrho}=(\mathbf{D}_{\varrho u})^{-1}(2\mathbf{A}_{u}+\partial_{\sigma}\mathbf{D}_{u\sigma})$,
$\varrho=1,2$, that is 
\begin{eqnarray}
\left[\begin{array}{c}
V_{1}\\
V_{2}
\end{array}\right] & = & -\frac{2}{\chi}\left[\begin{array}{c}
(\Gamma-\chi)/\alpha+\chi\beta+\lambda/\alpha^{2}\\
(\Gamma^{*}-\chi)/\beta+\chi\alpha+\lambda/\beta^{2}
\end{array}\right].
\end{eqnarray}
It is straightforward to verified that $\partial_{\beta}V_{1}=\partial_{\alpha}V_{2}=-2$
here. Thus, in the stationary limit $\dot{P}(\alpha,\beta,t=\infty)=0$,
we have 
\begin{eqnarray}
[\partial_{u}\mathbf{A}_{u}(\alpha,\beta)+\frac{1}{2}\partial_{u}\partial_{v}\mathbf{D}_{uv}(\alpha,\beta)]P(\alpha,\beta,t) & = & 0
\end{eqnarray}
or
\begin{equation}
\partial_{u}\{\mathbf{D}_{uv}[(\mathbf{D}_{vu})^{-1}(2\mathbf{A}_{u}+\partial_{v}\mathbf{D}_{uv})P+\partial_{v}P\}=0.
\end{equation}
The exact solution for the Fokker-Planck equation is therefore given
by
\begin{eqnarray}
P_{ss}(\alpha,\beta) & = & exp(-\int^{(\mathbb{C}_{\alpha},\mathbb{C}_{\beta})_{\varrho}}\mathbf{V}_{\varrho}(\alpha,\beta)d(\alpha,\beta)_{\varrho})\nonumber \\
 & = & exp\{\frac{2}{\chi}\int^{\mathbb{C}_{\alpha}}[(\Gamma-\chi)/\alpha+\chi\beta+\lambda/\alpha^{2}]d\alpha\nonumber \\
 &  & +\frac{2}{\chi}\int^{\mathbb{C_{\beta}}}[(\Gamma^{*}-\chi)/\beta+\chi\alpha+\lambda/\beta^{2}]d\beta\}\nonumber \\
 & = & exp\{\frac{2}{\chi}[(\Gamma-\chi)ln\alpha-\lambda/\alpha]+\frac{2}{\chi}[(\Gamma^{*}-\chi)ln\beta\nonumber \\
 &  & -\lambda/\beta]+2\int^{(\mathbb{C_{\alpha}},\mathbb{C}_{\beta})}(\alpha d\beta+\beta d\alpha)\}\nonumber \\
 & = & \alpha^{c-2}\beta^{d-2}exp[-z(1/\alpha+1/\beta)+2\alpha\beta],\nonumber \\
\end{eqnarray}
where $z=2\lambda/\chi$ and $c=2\Gamma/\chi$ ($d=c^{*}$) are dimensionless
quantities. Note that no Glauber-Sudarshan $P(\alpha,\alpha^{*})$
function exists in the steady state with $\beta=\alpha^{*}$ due to
the diverging exponential factor $\textrm{exp}(2\alpha\alpha^{*})$,
except as a generalized form $P(\alpha,\beta)$.

\begin{figure}
\begin{centering}
\includegraphics[width=0.9\columnwidth]{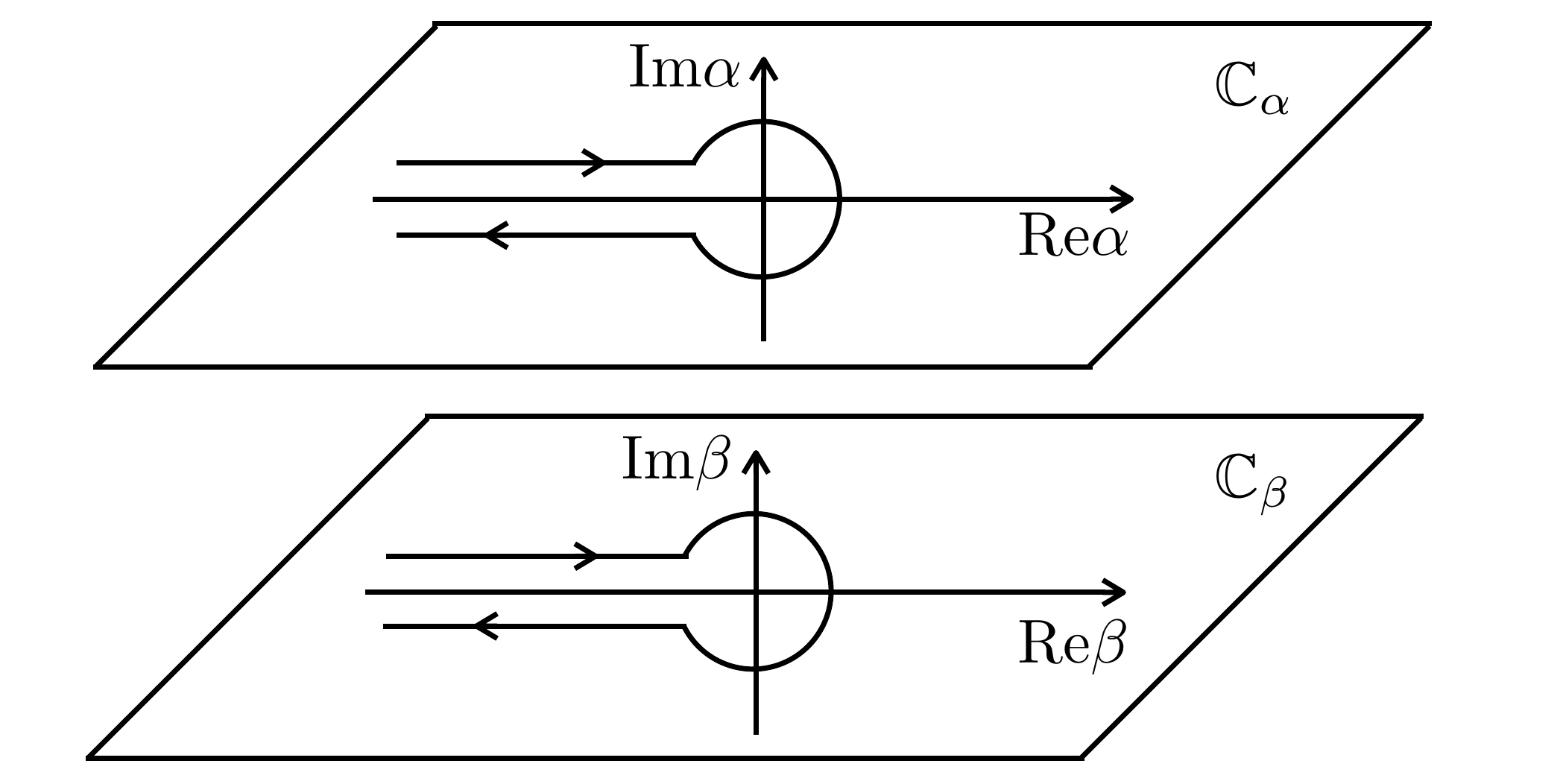}
\par\end{centering}

\caption{\label{fig:Hankel}The Hankel path of integration.}
\end{figure}

\begin{figure}
\begin{centering}
\includegraphics[width=1\columnwidth]{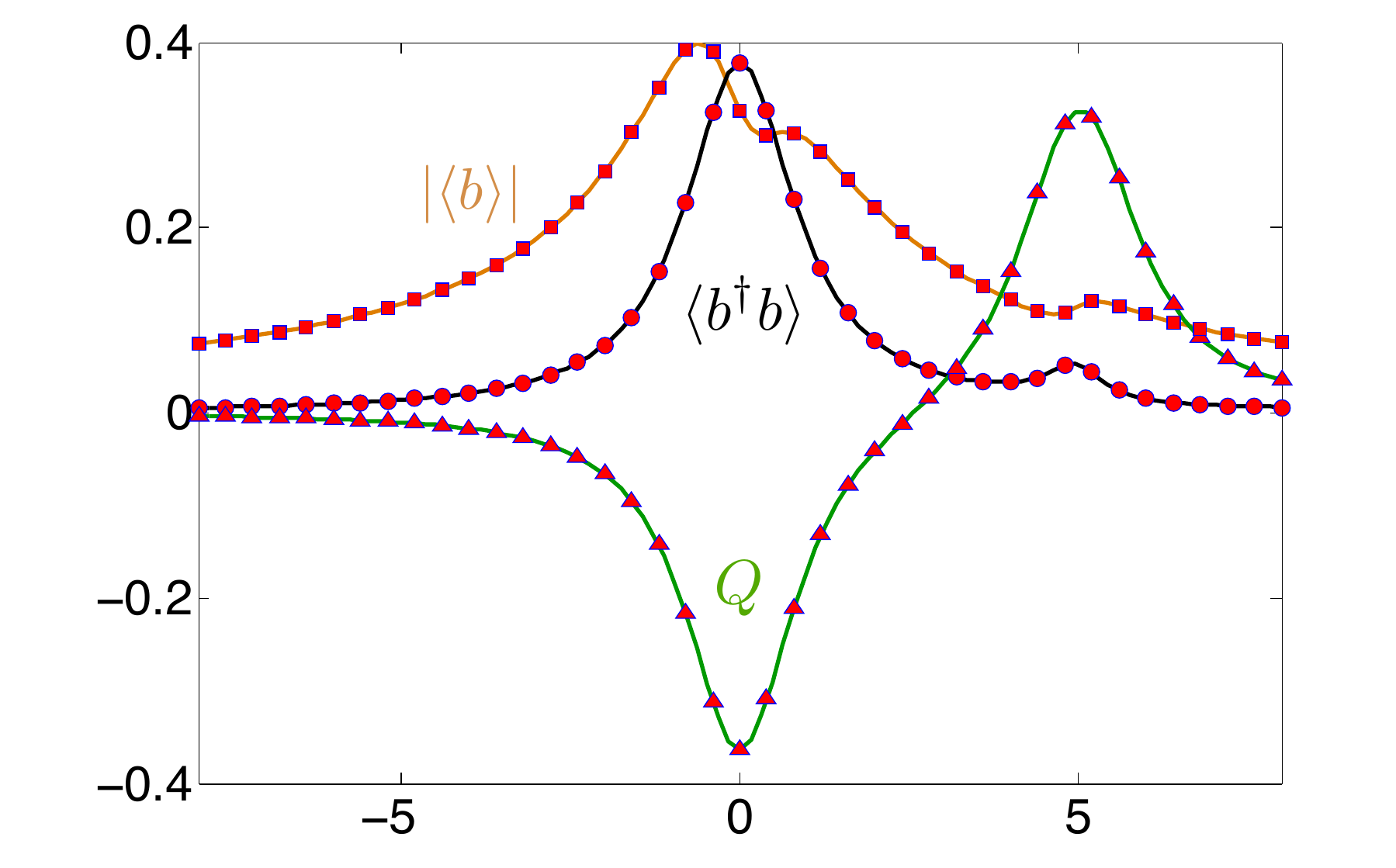}
\par\end{centering}
\caption{\label{fig_4s}(Color online) Comparison of the numerical steady state solution of $|\langle b\rangle|$, $\langle b^{\dagger}b\rangle$ and $Q$ under the original master equation for bosonization model (solid lines) with the analytical results Eqs. (\ref{eq:b_fp})-(\ref{eq:Q_fp}) (markers) in the main text. All parameters are the same as in figure \ref{fig: Phasediagram}(b).}
\end{figure}

\begin{figure}
\begin{centering}
\includegraphics[width=1\columnwidth]{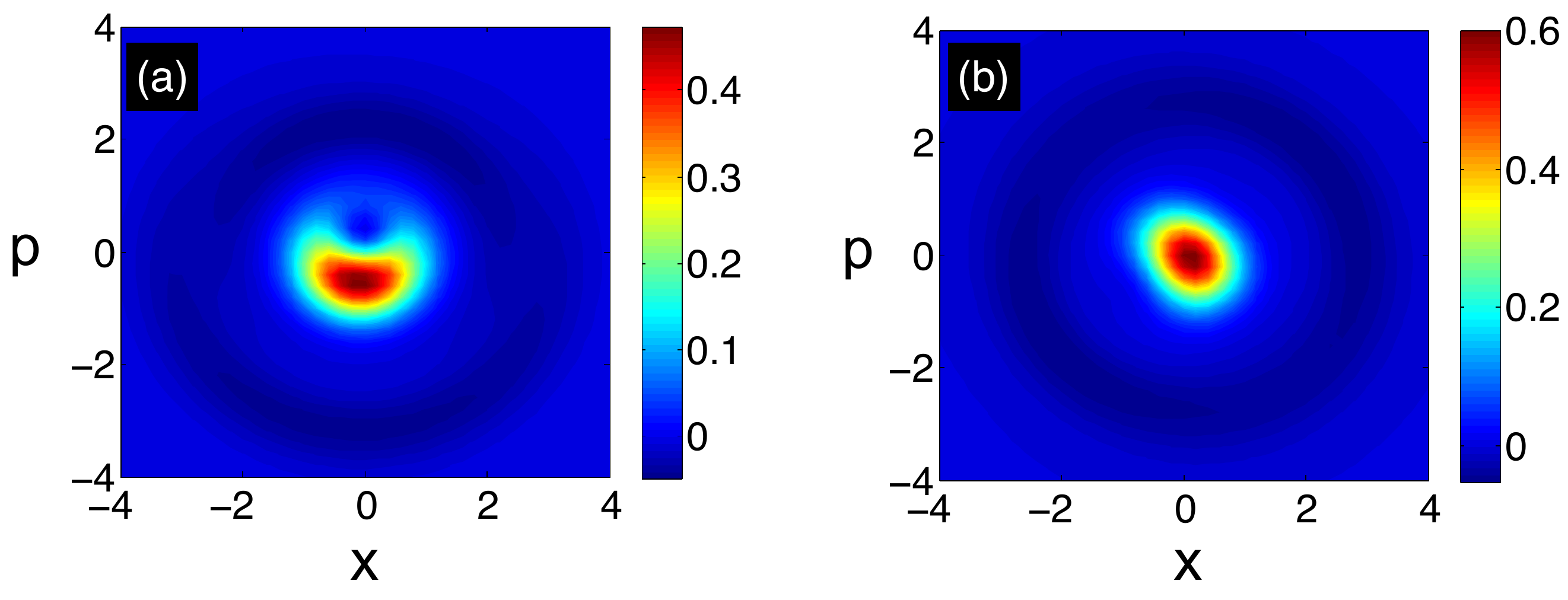}
\par\end{centering}
\caption{\label{fig_5s}(Color online) Wigner function $W(x,p)$ of the quantum mechanical
steady-state solution with (a) $\Delta=0$ and (b) $\Delta=\chi/2$.}
\end{figure}

The generalized $P(\alpha,\beta)$ function has to satisfying the
normalization condition, which implies
\begin{eqnarray}
I & = & \int^{(\mathbb{C}_{\alpha},\mathbb{C}_{\beta})}P_{ss}(\alpha,\beta)d\alpha d\beta\nonumber \\
 & = & \int^{(\mathbb{C}_{\alpha},\mathbb{C}_{\beta})}d\alpha d\beta\alpha^{c-2}\beta^{d-2}e^{-z(1/\alpha+1/\beta)+2\alpha\beta}\nonumber \\
 & = & \int^{(\mathbb{C}_{\alpha},\mathbb{C}_{\beta})}d(\alpha^{-1})d(\beta^{-1})\sum_{n=0}^{\infty}\frac{2^{n}}{n!}\alpha^{c+n}\beta^{d+n}e^{-z(1/\alpha+1/\beta)}\nonumber \\
 & = & \int^{(\mathbb{C}_{\alpha},\mathbb{C}_{\beta})}d\alpha_{1}d\beta_{1}\sum_{n=0}^{\infty}\frac{2^{n}}{n!}\alpha_{1}^{-(c+n)}\beta_{1}^{-(d+n)}e^{-z(\alpha_{1}+\beta_{1})}\nonumber \\
\end{eqnarray}
with $\alpha_{1}=\alpha^{-1}$ and $\beta_{1}=\beta^{-1}$. On the
other hand, the Gamma-function identity can be defined by using the
Hankel path of integration $\mathcal{O}$ (see Fig. \ref{fig:Hankel})
\begin{eqnarray}
\frac{1}{\Gamma(c+n)} & = & \frac{1}{2\pi i}\int_{\mathcal{O}}(\sigma z)^{-(c+n)}e^{\sigma z}d(\sigma z)\nonumber \\
 & = & \frac{z{}^{1-(c+n)}}{2\pi i}\int_{\mathcal{O}}\sigma{}^{-(c+n)}e^{\sigma z}d\sigma,
\end{eqnarray}
with which we obtain
\begin{eqnarray}
I & = & \int^{(\mathbb{C}_{\alpha},\mathbb{C}_{\beta})}d\alpha_{1}d\beta_{1}\sum_{n=0}^{\infty}\frac{2^{n}}{n!}\alpha_{1}^{-(c+n)}\beta_{1}^{-(d+n)}e^{-z(\alpha_{1}+\beta_{1})}\nonumber \\
 & = & \sum_{n=0}^{\infty}\frac{2^{n}}{n!}[\int_{\mathcal{O}}d\alpha_{1}\alpha_{1}^{-(c+n)}e^{-z\alpha_{1}}\int_{\mathcal{O}}d\beta_{1}\beta_{1}^{-(d+n)}e^{-z\beta_{1}}]\nonumber \\
 & = & (-4\pi^{2})\sum_{n=0}^{\infty}\frac{2^{n}}{n!}\frac{z{}^{c+d+2(n-1)}}{\Gamma(c+n)\Gamma(d+n)}\nonumber \\
 & = & \frac{(-4\pi^{2})z{}^{c+d-2}}{\Gamma(c)\Gamma(d)}{}_{0}F_{2}(c,d,2z^{2}),
\end{eqnarray}
where $_{0}F_{2}(c,d,z)$ is the generalized hypergeometric function
defined as
\begin{equation}
_{0}F_{2}(c,d,z)=\sum_{n=0}^{\infty}\frac{z^{n}}{n!}\frac{\Gamma(c)\Gamma(d)}{\Gamma(c+n)\Gamma(d+n)}.
\end{equation}

The normalized $i$th-order correlation function $G^{(i,j)}$ corresponding
to the normally ordered averages $\langle b^{\dagger i}b^{j}\rangle$
can be calculated by the generalized $P$-representation:
\begin{eqnarray}
G^{(i,j)} & = & \frac{\int d\sigma(\alpha,\beta)P(\alpha,\beta)\beta^{m}\alpha^{n}}{\int d\sigma(\alpha,\beta)P(\alpha,\beta)}\nonumber \\
 & = & \frac{\frac{(-4\pi^{2})z{}^{c+d-2+i+j}}{\Gamma(c+j)\Gamma(d+i)}{}_{0}F_{2}(c+j,d+i,2z^{2})}{\frac{(-4\pi^{2})z{}^{c+d-2}}{\Gamma(c)\Gamma(d)}{}_{0}F_{2}(c,d,2z^{2})}.\nonumber \\
\end{eqnarray}
With the Euler's functional equation $\Gamma(z)/\Gamma(z+n+1)=1/z(z+1)\cdots(z+n)$,
we finally have
\begin{equation}
G^{(i,j)}=z{}^{i+j}\frac{\Gamma(c)\Gamma(d){}_{0}F_{2}(c+j,d+i,2z^{2})}{\Gamma(c+j)\Gamma(d+i){}_{0}F_{2}(c,d,2z^{2})},
\end{equation}
and therefore the quantum mechanical steady-state solution of $\langle b\rangle$,
$\langle b^{\dagger}b\rangle$ and $Q=\frac{\langle b^{\dagger}b^{\dagger}bb\rangle}{\langle b^{\dagger}b\rangle}-\langle b^{\dagger}b\rangle$.  We have verified these analytical results by comparisons with the numerically obtained steady state solution of the original master equation $\dot{\rho}=-i[H_{b},\rho]+\frac{\gamma}{2}(2b\rho b^{\dagger}-\rho b^{\dagger}b-b^{\dagger}b\rho)$. And as shown in figure \ref{fig_4s}, we find excellent agreement between the two different ways.

\section{\label{apd_D}The nonclassical nature of the quantum mechanical steady
states}

The nonclassical nature of the quantum mechanical steady states for
the excitation processes with resonant driving and the laser detuning
$\Delta=\chi/2$ manifests itself in the Wigner function with negative
value, as shown in figure \ref{fig_5s}.

\bibliographystyle{apsrev}

\bibliography{ref2,Rydberg_atom,Rydberg_atom_2}
\end{document}